\RequirePackage[l2tabu, orthodox]{nag}
\RequirePackage{snapshot}

\documentclass[9pt,onecolumn]{extarticle}
\makeatletter
\if@twocolumn
  \usepackage[dvips,letterpaper,top=0.5in, bottom=0.5in, left=0.75in, right=0.5in,includefoot,heightrounded]{geometry}
\else
  \usepackage[dvips,letterpaper,margin=1in,includefoot,heightrounded]{geometry}
\fi

\usepackage{srcltx}

\usepackage[russian,portuges,english]{babel}

\iflanguage{portuges}
    {\newcommand{\keywordname}{Palavras-chaves}}
    {\newcommand{\keywordname}{Keywords}}

\usepackage{amsmath}
\usepackage{amssymb,amsfonts}

\usepackage{abstract}

\usepackage{graphicx}
\usepackage[usenames,dvipsnames,svgnames,x11names]{xcolor}
\usepackage{subfigure}
\usepackage{psfrag}

\usepackage{booktabs}

\usepackage{setspace}
\usepackage{flushend}

\usepackage{cite}

\usepackage{hyperref}
\usepackage[normalem]{ulem}

\usepackage{enumerate}

\usepackage[short,12hr]{datetime}

\newtheorem{theorem}{Theorem}

\makeatletter

\newcommand{\printtitle}{%
\makeatletter
\if@twocolumn

\twocolumn[%
  \maketitle
  \begin{onecolabstract}
    \myabstract
  \end{onecolabstract}
  \begin{center}
    \small
    \textbf{\keywordname}
    \\\medskip
    \mykeywords
  \end{center}
  \bigskip
]
\saythanks
\else
  \maketitle
  \begin{onecolabstract}
    \myabstract
  \end{onecolabstract}
  \begin{center}
    \small
    \textbf{\keywordname}
    \\\medskip
    \mykeywords
  \end{center}
  \bigskip
  \onehalfspacing
\fi
\makeatother
}

\author{%
A. Borges Jr.%
\thanks{%
A. Borges Jr.
was with the
Universidade Federal de Pernambuco, Brazil.}
\and
R. J. Cintra%
\thanks{%
R. J. Cintra is with the
Signal Processing Group,
Universidade Federal de Pernambuco, Brazil.
E-mail: \url{rjdsc@de.ufpe.br}}
\and
D. F. G. Coelho%
\thanks{D. F. G. Coelho is an independent researcher, Calgary, Canada.
E-mail: \url{diegofgcoelho@gmail.com}}
\and V. S. Dimitrov%
\thanks{%
V. S. Dimitrov is with the
University of Calgary, Canada.}
}

\title{%
Low-complexity Architecture for
AR(1) Inference}

\newcommand{\myabstract}{%
In this Letter, we propose
a low-complexity estimator for
the correlation coefficient
based on the signed $\operatorname{AR}(1)$ process.
The introduced
approximation
is suitable for
implementation in low-power hardware
architectures.
Monte Carlo simulations
reveal that the
proposed estimator
performs comparably to
the competing methods in literature
with maximum error in order of $10^{-2}$.
However,
the hardware implementation
of the introduced method presents
considerable advantages
in several relevant metrics,
offering more than 95\% reduction in dynamic power
and
doubling the maximum operating frequency
when compared to the reference method.
}

\newcommand{\mykeywords}{%
AR(1) inference,
low-complexity algorithms
}

\date{}

\begin{document}

\printtitle

\section{Introduction}

Due to the raising demand
for digital signal processing (DSP) systems
capable of operating
at
low power
and
low complexity,
approximate methods
have been considered
for image processing~\cite{haweel2001,cintrabayer}.
In particular,
several approximate discrete transforms
have been recently proposed~\cite{Tablada2015, deA.Coutinho2017, Zhao2016}
for image compression,
where pixel data often stems from natural images
and are modeled according to
the first order autoregressive ($\operatorname{AR(1)}$) process~\cite{rao2014}.
The $\operatorname{AR}(1)$ model depends
only on a single parameter, the correlation coefficient~$\rho$,
whose identification
determines the suitable DSP tools for data analysis~\cite{Britanak2007, Pourazad2012}.
In particular,
image sensor networks
and
mobile computing systems
may benefit from low-complexity
fundamental DSP building blocks~\cite{betzel2018}.
In this Letter,
we aim at the derivation of
a low-complexity algorithm
for the estimation
of $\rho$
targeting its implementation on embedded,
low-power devices.

\section{$\operatorname{AR}(1)$ Processes}

A real-valued, discrete-time, wide-sense stationary
stochastic process $\{X_n, n=1,2,\ldots\}$
with null mean and finite variance is said
to be an $\operatorname{AR}(1)$ process if
\begin{equation}
\label{eq:markov}
X_n = \rho \ X_{n-1} + W_n
,
\end{equation}
where
$|\rho| < 1$
and
$W_n$ is a white noise
process independent of $X_n$.
If the joint distribution of
any finite set of samples
from~\eqref{eq:markov}
is Gaussian,
then we say that $X_n$ is
a Gaussian $\operatorname{AR(1)}$ process.
Assuming stationarity of $X_n$,
its autocorrelation function
is given by $\mathrm{cor}(X_n, X_m) = \rho^{|n-m|}$,
which depends solely on $\rho$.
The traditional estimator for~$\rho$
is given by~\cite[p.~77]{Kay1998}:
\begin{equation}
\label{eq:exact_estimator}
\hat{\rho}_N
=
\frac
{\sum_{n=2}^N X_n X_{n-1}}
{\sum_{n=1}^N X_n^2}
.
\end{equation}
Hereafter
we refer to the above statistic
as the autocorrelation function (ACF) estimator~\cite{Kay1998}.

\section{Signed $\operatorname{AR}(1)$ Processes}

A binary threshold process derived from $X_n$
is defined according
to
$S_n = I(X_n > 0)$,
where $I(\cdot)$ equals $1$ if its
argument is true and $0$ otherwise.
We refer to $S_n$ as the signed~$\operatorname{AR}(1)$ process.
The following theorem proposed by Kedem~\cite{kedem1980}
relates $\rho$ to the stochastic structure of $S_n$.
\begin{theorem}%
\label{thm:markov}
If $X_n$ is a Gaussian $\operatorname{AR(1)}$ process,
then the signed $\operatorname{AR(1)}$ $S_n$ is a Markov chain
over states $\{0,1\}$
with symmetric transition probabilities matrix.
Let $\lambda$ be the probability of remaining at
the same state.
Then the correlation coefficient $\rho$ is given by
\begin{equation}
\label{eq:rho_lambda}
\rho(\lambda)
=
\cos
\Big(
\pi (1-\lambda)
\Big)
.
\end{equation}
\end{theorem}
The unbiased maximum likelihood estimator (MLE) of~$\lambda$
is
based on~$S_1, S_2,\ldots,S_{N}$
and is given by
\begin{equation}
\label{eq:lambda}
\hat{\lambda}_N
=
\frac{1}{N-1} \sum_{n=2}^N I(S_{n} = S_{n-1})
.
\end{equation}
Invoking the
Invariance Principle~\cite{casellaberger},
the MLE for~$\rho$ is derived from~\eqref{eq:rho_lambda}
and
is given by $\rho(\hat{\lambda}_N)$.
We refer to it as the Kedem estimator.

\section{Approximate Estimation}

A low-complexity estimator for $\rho$
can be derived by
means of approximating the function~$\rho(\lambda)$ in~\eqref{eq:rho_lambda}.
By using global and convex optimization~\cite{Storn1997, pwlf},
we obtain
the optimal 5-interval
piecewise linear approximation for~$\rho(\lambda)$ in~\eqref{eq:rho_lambda},
furnishing
the following proposed low-complexity estimator:
\begin{equation}
\tilde{\rho}(\lambda) = \begin{cases}
 -1.01+0.64\lambda, & \mbox{if } \lambda \in [0.00,0.14), \\
 -1.20+1.97\lambda, & \mbox{if } \lambda \in [0.14,0.30), \\
 -1.51+3.02\lambda, & \mbox{if } \lambda \in [0.30,0.70), \\
 -0.77+1.97\lambda, & \mbox{if } \lambda \in [0.70,0.86), \\
\phantom{-}0.37+0.64\lambda, & \mbox{if } \lambda \in [0.86,1.00). \\
\end{cases}
\label{eq:rho_tilde}
\end{equation}
The absolute error satisfies:
$|\tilde{\rho}(\lambda)-\rho(\lambda)| < 1.4\cdot 10^{-2}$.
The proposed approximate estimator is therefore
$
\tilde{\rho}(\hat{\lambda}_N)
$.

\section{Simulations}

\begin{figure}

 \centering
 \psfrag{p}[][][1.0]{$\rho$}
 \psfrag{Bias}[][][1.0]{Bias}
 \psfrag{ACFXXXXX}[][][1.0]{$\operatorname{err}(\hat{\rho}_N)$}
 \psfrag{KEDEMXXX}[][][1.0]{$\operatorname{err}({\rho}(\hat{\lambda}_N))$}
 \psfrag{PROPXXXX}[][][1.0]{$\operatorname{err}(\tilde{\rho}(\hat{\lambda}_N))$}
 \psfrag{-3e-2XX}[][][0.8]{$-3\cdot 10^{-2}$}
 \psfrag{-2e-2XX}[][][0.8]{$-2\cdot 10^{-2}$}
 \psfrag{-1e-2XX}[][][0.8]{$-1\cdot 10^{-2}$}
 \psfrag{0}[][][0.8]{$0$}
 \psfrag{1e-2XX}[][][0.8]{$1\cdot 10^{-2}$}
 \psfrag{2e-2XX}[][][0.8]{$2\cdot 10^{-2}$}
 \psfrag{3e-2XX}[][][0.8]{$3\cdot 10^{-2}$}
\subfigure[\fontsize{8}{8}\selectfont Average bias for ACF estimator.\label{fig:ACF}]{\includegraphics{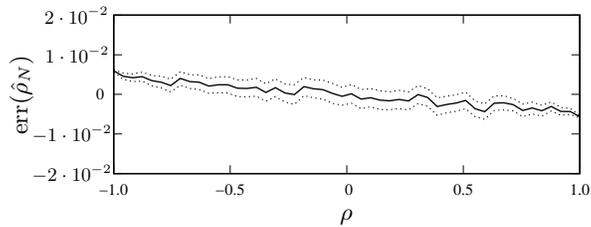}}

\subfigure[\fontsize{8}{8}\selectfont Average bias for Kedem estimator.\label{fig:Kedem}]{\includegraphics{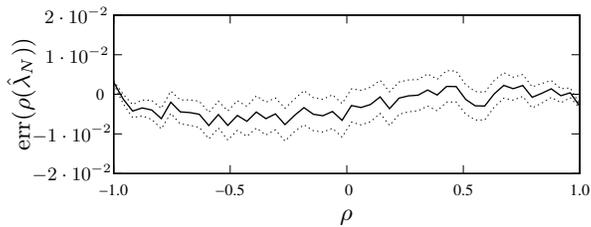}}

\subfigure[\fontsize{8}{8}\selectfont Average bias for proposed estimator.\label{fig:prop}]{\includegraphics{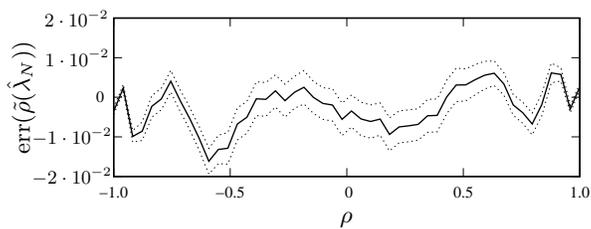}}

\caption{Average bias for
for the ACF estimator~(\ref{fig:ACF}),
Kedem estimator~(\ref{fig:Kedem}),
and
the proposed approximated estimator~(\ref{fig:prop}),
respectively,
over an ensemble of $1000$ Monte Carlo replicates.
The upper and bottom dotted lines for each
plot represent the~95\% confidence interval based on the normal distribution.}
\label{fig:rho_mc_Gaussian}
\end{figure}

A Monte Carlo simulation with~$R = 1000$ replicates
of the~$\operatorname{AR}(1)$ process of length $N = 512$
was used to assess
behavior of the proposed estimator
in comparison with the ACF estimator.
The selected values of~$\rho$
were $\rho \in [-1, 1]$ in steps of~$4\cdot 10^{-2}$.
The power of the additive white noise
in~\eqref{eq:markov}
was adjusted
such that~$90\%$ of its realizations
are within~$[-1, 1]$,
resulting in~$W_n \sim \mathcal{N}(0, 0.61)$.
In order to quantify
the performance of the proposed
low complexity estimator
compared to the ACF estimator in~\eqref{eq:exact_estimator},
we computed
$\hat{\rho}_N^{(r)} - \rho$,
$\rho^{(r)}(\hat{\lambda}_N) - \rho$,
and
$\tilde{\rho}_N^{(r)}(\hat{\lambda}_N) - \rho$
for each replicate~$r = 0, 1, \ldots, R-1$,
where
$\hat{\rho}_N^{(r)}$,
$\rho^{(r)}(\hat{\lambda}_N)$,
and
$\tilde{\rho}^{(r)}(\hat{\lambda}_N)$
are the
estimates
of~$\rho$ according to
the ACF,
Kedem,
and the proposed estimators
for each  replicate~$r$,
respectively.
Fig.~\ref{fig:rho_mc_Gaussian} displays values
of
$
\operatorname{err}(\hat{\rho}_N) =
\sum_{r=1}^R
(\hat{\rho}_N^{(r)} - \rho) / R
$,
$
\operatorname{err}(\rho(\hat{\lambda}_N)) =
\sum_{r=1}^R
(\rho^{(r)}(\hat{\lambda}_N) - \rho) / R
$,
and
$
\operatorname{err}(\tilde{\rho}(\hat{\lambda}_N)) =
\sum_{r=1}^R
(\tilde{\rho}^{(r)}(\hat{\lambda}_N) - \rho) / R
$
along with the~$95\%$ confidence intervals
based on the normal distribution.

\section{Arithmetic Complexity}

The direct implementation of the ACF estimator
requires
one division,
$2N$ multiplications,
$2N-1$ additions.
Assuming that $\hat{\lambda}_N$ is known,
the estimator for $\rho$
derived from Theorem~\ref{thm:markov}
requires
one addition,
one multiplication by $\pi$,
and one
call of the cosine function.
The cosine function is implemented through
the coordinated rotation digital computer
(CORDIC) algorithm,
which is an iterative method that
employs successive additions of
bit-shifted quantities.
Each iteration of the CORDIC algorithm
requires two additions and two shifts at most,
depending on the the angles that are computed.
The implementation of
the CORDIC block
used in the the proposed architectures
require 14~iterations~\cite[pg.~40]{xilinx-cordic},
resulting in a total of 28 additions and up to 28 bit-shifts
for each evaluation of the cosine function.
On its turn,
the computation
of~$\hat{\lambda}_N$
requires
$N-1$ additions,
$N-1$ comparisons,
and
one division by~$N-1$.
On the other hand,
given that~$\hat{\lambda}_N$ is available,
the proposed approximate estimator
$\tilde{\rho}(\hat{\lambda}_N)$
based on~\eqref{eq:rho_tilde}
requires
only multiplications by simple constants,
additions and bit-shifting operations,
amounting to~$N+1$ additions and at most~$2$ shifts.
as well as a comparator.
Table~\ref{tab:arithmetic-complexity}
summarizes the arithmetic complexity
for the ACF, Kedem, and the proposed estimator.

\begin{table}
\centering
\caption{Complexity arithmetic summary for the implemented designs}
\label{tab:arithmetic-complexity}
\begin{tabular}{ccccc}
\toprule
Estimator & Multiplication & Division & Addition & Shifts \\
\midrule
ACF & $2N$ & $1$ & $2(N-1)$ & $0$\\
\midrule
Kedem & $1$ & $1$ & $N+28$ & $28$\\
\midrule
Proposed & $0$ & $0$ & $N+1$ & $2$\\
\bottomrule
\end{tabular}

\end{table}

\section{Hardware Implementation}

\begin{figure}
 \centering
 \includegraphics[scale=1]{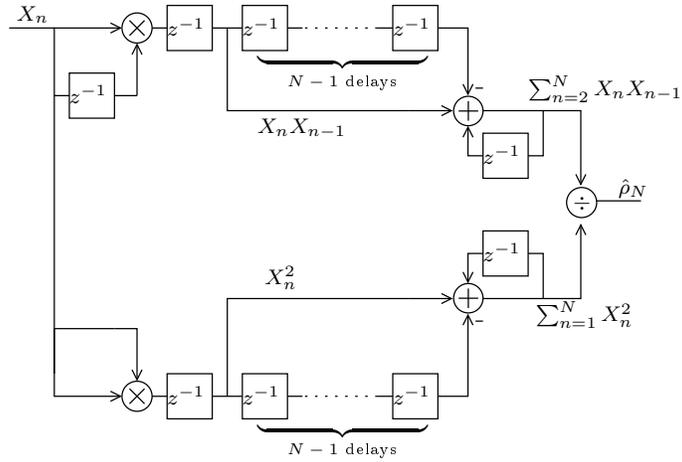}
 \caption{Architecture for the ACF estimator.}
 \label{fig:correlator}
\end{figure}

The ACF, Kedem, and the proposed approximate estimator
were implemented
on a
Xilinx Artix-7 XC7A35T-1CPG236C FPGA device.
Although there are different architectures for digital correlators based on fast Fourier transforms (FFT)
for different applications~\cite{Balu2017}, we choose not to implement such scheme
given its higher complexity in terms of resources and power compared to the architecture in~Fig.~\ref{fig:correlator}.
The implementations are capable of
providing an estimate of~$\rho$ for every clock pulse
using
the last~$N$ input samples.
The ACF estimator in~\eqref{eq:exact_estimator} was
implemented using the architecture depicted in Fig.~\ref{fig:correlator}.
The structure for computing
$\hat{\rho}_N$
possesses
two identical $(N-1)$-sample delay lines
after
the computation of
$X_nX_{n-1}$ and~$X_nX_{n}$.
The values of $X_nX_{n-1}$ and~$X_nX_{n}$
in~$N-1$ cycles in the past
are subtracted
from the current value of
$\sum_{n=2}^{N}X_nX_{n-1}$ and~$\sum_{n=1}^{N}X_nX_{n}$, respectively.
This scheme allows the overall system to compute the correlation
limited to the last~$N$ samples.
Without the delay network,
the circuit would yield the correlation for
the whole sequence from its beginning
and incur in overflow.

\begin{figure}
 \centering
 \includegraphics[scale=1]{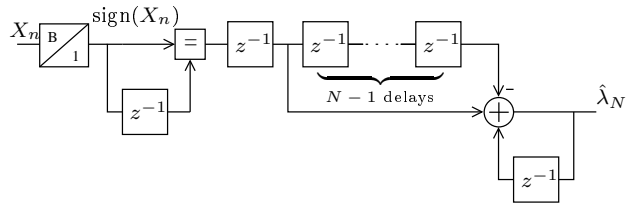}
 \caption{Architecture for the estimation of $\lambda$.}
 \label{fig:low-comp-correlator}
\end{figure}

The Kedem and the proposed estimators
share the structure shown in Fig~\ref{fig:low-comp-correlator},
which is required for computing
the estimated value of $\lambda$.
The input word representing the samples in time in~Fig.~\ref{fig:low-comp-correlator} is downsized from~$B = 10$ bits to the sign bit.
The current sign bit is then compared to the sign bit of the last sample
and the result
is stored in a shift register of size~$N$.
We adopted $N = 512$.
Subsequently
the output sequence from the comparator
is
shifted to the right at every rising edge of the clock.
The value of~$\hat{\lambda}$ is then added to the net value of the immediate sign bit comparison and the
comparison~$N$ clock periods earlier.
Note that this delay network
is the same present on Fig.~\ref{fig:correlator},
allowing the design to account for only a window of size~$N$.
This negative loop results in forcing~$\hat{\lambda}$
to store the number of comparisons
that were
evaluated as true in the last~$N$ clock pulses,
including the current sample compared to the previous one.

\begin{figure}
 \centering
 \includegraphics[scale=1]{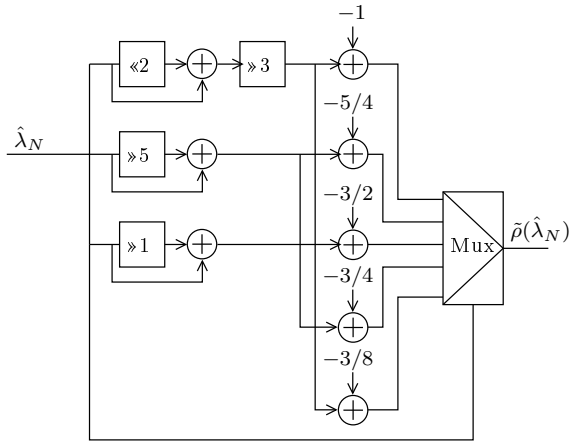}
 \caption{Architecture for the implementation of~$\tilde{\rho}(\lambda)$,
where the constants in~\eqref{eq:rho_tilde}
where approximated by the ones in Table~\ref{tab:constants}.
}
\label{fig:rho_tilde}
\end{figure}

\begin{table}
\centering

\caption{Dyadic approximation for the constants required in~\eqref{eq:rho_tilde}}
\label{tab:constants}

\begin{tabular}{ccccccccc}
\toprule
Const.&$0.64$ & $1.97$ & $3.02$ & $1.01$ & $1.20$ & $1.51$ & $0.77$ & $0.37$\\
\midrule
Approx.&\!\!\!$5/8$ &\!\!\!$63/32$\!\!\!&\!\!\!$3$\!\!\!&\!\!\!$1$\!\!\!&\!\!\!$5/4$\!\!\!&\!\!\!$3/2$\!\!\!&\!\!\!$3/4$\!\!\!&\!\!\!$3/8$\!\!\!\\
\bottomrule
\end{tabular}

\end{table}

\begin{table}
\centering

\caption{Resource utilization for FPGA implementation using 10-bit wordlength}
\label{tab:hardware-implementation}

\begin{tabular}{rrrr}
\toprule
Resource & ACF~\cite{Kay1998} & Kedem~\cite{kedem1980} & Proposed \\%& Variation\\
\midrule
LUT & 1365 & 437 ($-$67.98\%) & 98 ($-$93\%)\\
\midrule
FF & 2762 & 853 ($-$69.11\%) & 582  ($-$79\%)\\
\midrule
Slices & 513 & 251 ($-$51.07\%) & 129  ($-$75\%)\\
\midrule
$f_\text{max}$ (MHz) & 115.64 & 165.75 ($+$30.22\%) &242.30 ($+$110\%)\\
\midrule
Latency (cycles) &35 & 22 ($-$37.14\%) & 2  ($-$94\%)\\
\midrule
Power (mW) &76& 6 ($-$92.10\%) & 4  ($-$95\%)\\
\bottomrule
\end{tabular}

\end{table}

Such estimated value $\hat{\lambda}_N$
is then submitted
to the
block implementing the
functions
$\rho(\cdot)$
or
$\tilde{\rho}(\cdot)$
for the Kedem
or the proposed estimator,
respectively.
For the Kedem estimator,
a single call for the cosine function
is required,
being physically implemented
according to the Xilinx
implementation of
the
CORDIC algorithm
as described
in~\cite[p.~17]{xilinx-cordic}.

In terms of the proposed estimator,
aiming at a low-cost hardware implementation,
the constants
required in~\eqref{eq:rho_tilde}
were
approximated
according to
dyadic integers with low magnitude numerator,
rendering the values in Table~\ref{tab:constants}.
The computation of~$\tilde{\rho}(\lambda)$
is depicted in Fig.~\ref{fig:rho_tilde}.
Pipeline stages are not shown for simplicity.
The mux block in Fig.~\ref{fig:rho_tilde}
selects the appropriate path according to
the interval defined in~\eqref{eq:rho_tilde}.
The symmetry of~$\tilde{\rho}(\lambda)$
was exploited
in such a way that
the slope coefficients
in the first three intervals
were sufficient for the computing~$\tilde{\rho}(\lambda)$
for all possible values of~$\lambda$.

The designs were implemented using a signed 10-bit word
for representing the the output estimates
according to the ACF, Kedem, and the proposed estimator.
Table~\ref{tab:hardware-implementation}
summarizes
the resource utilization
in terms of look-up table (LUT),
flip-flops (FF), and slices~\cite{slice-xilinx}
and
performance measurements
expressed
by
maximum operating frequency,
latency,
and
dynamic power.
The percentages in parenthesis
inform the variations compared
to the ACF estimator.
The
proposed design offers significant savings
in resource consumption:
(i)~the number of LUTs, latency, and dynamic power
were dramatically reduced in more than 93\%
compared
to the exact implementation of the estimator.
The number of FFs and slices were reduced in more than 75\%;
and
the maximum operating frequency
received
a two-fold increase.
The significant reduction in the latency of the design based on~\eqref{eq:rho_lambda}
is mainly due to the absence of
multipliers and dividers~\cite{xilinx-divider},
which demand several clock cycles to complete an operation.
In particular, the proposed design shows better metrics
because the consecutive shift-and-add operations of the CORDIC~\cite{xilinx-cordic} block,
employed to compute the cosine,
are substituted by
multiplications by hardware-friendly constants, requiring just
a few shifts and additions.

\section{Conclusions}

A low-complexity approximate method for
computing
the correlation coefficient
in $\operatorname{AR}(1)$ processes
was introduced.
Numerical simulations
indicate the good performance
of the proposed estimator
when compared
with the standard method in literature.
The associate
computational complexity favors its
implementation in low-power hardware.
Hardware implementation metrics of the proposed
estimator are shown to be much more attractive
than the ones
resulting from the ACF estimator architecture.
In particular, the dynamic power of the implementation of
the proposed method is almost fourteen times smaller than the traditional method,
while the the maximum operating frequency is doubled.

\section*{Acknowledgments}

This work has been partially supported by CNPq, Brazil, and NSERC, Canada.

{\small
\singlespacing
\bibliographystyle{siam}
\bibliography{refs}
}

\end{document}